\documentclass[12pt,twocolumn]{article}
\usepackage{array}
\usepackage{amssymb}
\usepackage[table]{xcolor}
\usepackage{color}
\usepackage{graphicx}
\usepackage{amsmath}
\usepackage{cases}
\usepackage{multirow}
\usepackage{hyperref}
\usepackage[margin=0.9in]{geometry}

\makeatletter
\begin{document}

\title{Coinbugs: Enumerating Common Blockchain Implementation-Level Vulnerabilities}
\author{Aleksandar Kircanski and Terence Tarvis\\NCC Group}
\date{}
\maketitle
\begin{abstract}
A good amount of effort has been dedicated to surveying and systematizing Ethereum smart contract security bug classes, see e.g. \cite{DASP, smartcontract,trailofbits}. There is, however, a gap in literature when it comes to surveying implementation-level security bugs that commonly occur in basic PoW blockchain node implementations, discovered during the first decade of Bitcoin's existence. This paper attempts to fill this void. In particular, if software which participates in a network by validating and generating new blocks is developed from scratch, WCGW - What Could Go Wrong?

Ten broad bug type categories are listed and for each category, known examples are linked. Blockchain, as designed by the  Satoshi's paper is exciting and introduces several novel bug classes which are interesting to security researchers. The paper is aimed at security testers aiming to start out in blockchain security reviews and blockchain developers as a reference on common pitfalls.
\end{abstract} 

\section{Introduction}
Over a period of more than a decade, the Bitcoin community invested a large amount of
effort in developing and analyzing the Bitcoin blockchain implementation. A number of
consensus, cryptographic, denial of service and application level type of issues have
been uncovered, as shown by the Bitcoin CVE list \cite{bitcoincve}. 

While some of this information is captured by the Bitcoin CVE list, security testers
and blockchain developers still lack a resource that helps fast ramp up when it comes to
security bug types. In this paper we attempt to close that gap and help security researchers
extrapolate the issues to implementations beyond Bitcoin. In particular, while many of the issues
were first identified in Bitcoin, they are archetypical for blockchain and re-occur
in different blockchain contexts and flavors. 

This paper assumes basic understanding of blockchain and provides a top 10 style bug class list,
with comments and references to examples. For those unfamiliar with blockchain and starting out
in this area, good initial references can be found in the Bitcoin developer guide and a free book \cite{masteringbitcoin}.
After reading the materials linked to in the sections below, a researcher should be
well-equipped to start auditing non-Bitcoin blockchain software (though the end
result may be bugs outside of this paper's list). 

Smart contract level blockchain issues are excluded from the scope of this paper, since the scope are
the basic  blockchain implementation security issues. The following bug types are discussed in the sections below:
\textbf{\begin{enumerate}
\item Netsplit due to multiple client implementations
\item Netsplit due to execution environment discrepancies
\item Netsplit via block hash poisoning
\item Netsplit via unintended or pre-mature fork
\item Netsplit via branch confusion
\item Improper timestamp validation
\item Integer underflow/overflow
\item Merkle tree implementation issues
\item Storage exhaustion in block or transaction processing
\item CPU exhaustion in block or transaction processing
\end{enumerate}}
An additional section on other blockchain client implementation issues is also at the end of the paper.

It is expected that in future more blockchain implementations will operate bug bounty programs
and we're hoping this paper will be a good reference for bug bounty hunters. While most of the
issues listed here concern Proof-of-Work (PoW) systems systems such as Bitcoin and Ethereum,
it is likely that variants of these bugs plague non-PoW and various variants of specific consensus
implementations too. 

As mentioned, blockchain, as designed by the Satoshi's paper is exciting and introduces several novel
bug classes which are interesting to security researchers. One of the most interesting bug classes in blockchain
are netsplit attacks, discussed in the first section below.
\subsection{Netsplit}
From a security tester's perspective, blockchain is exciting: it is brittle in a way most of the software out there is not. 
This brittleness comes from a blockchain's security requirement that the clients acting on the network behave the exact same
way when it comes to validating blocks. For example, there should not exist a block with a special quirk that's always rejected 
by subset of nodes, but can be validated by the remaining nodes. Any such quirk needs to either get a block rejected on all of 
the nodes, or have no impact on the validation/rejection decision. This is very challenging given the high complexity involved.

A netsplit condition occurs if a subset of the nodes cannot validate a block the remaining nodes accept, which results in existence
of two blockchains in parallel.  While splitting a network this way may be thought of as a Denial of Service (DoS) attack, it is important 
to stress that the netsplit condition facilitates a double-spend attack (in which a single coin is spent more than once). Once the 
network unifies, transactions in one of the two blockchains will be reverted. If an attacker manages to prevent these transactions 
from being replayed on or reaching the original chain, a  double-spend attack may happen. As such, mitigating netsplit attack vectors 
is of highest importance in the blockchain world. 

Now it may seem that this brittleness caused by subtle differences in library implementations
is unique to blockchain and that only blockchain bug hunters can profit from it. This, however, 
is not the case, differences in library implementations have found to be a source of security problems in the past. 
Suppose there exist multiple implementations of a given protocol. These different implementations are likely to in fact implement 
their own dialects of the protocol, and not be bug-for-bug compatible. Does the mere existence of such innocuous discrepancies result 
in security problems? See these examples:

\begin{enumerate}
\item TCP/IP packet processing discrepancies behind insertion and evasion attacks on IDS systems \cite{ptacek}:  Assume an Intrusion Detection System  (IDS) listening on the network for attack patterns. It is processing packets, just as the attacker's target. However, the two packet processors may be subtly different in what packets they considered valid. If an attacker can craft packets that will be rejected by the target host and accepted by the IDS (or vice versa), the view of the detection system will be distorted, failing its mission. 
\item BGP protocol implementation differences issues \cite{bgp}: Sending an odd BGP packet may cause a part of the network to accept and forward it and the other part of the network to drop connections to all routers it received it from, causing a BGP level confusion.
\item Certificate validation bypass via X.509 processing discrepancies \cite{x509}: Consider a CA and a browser leveraging two different certificate processing libraries,  with known discrepancies. When the CA validates a certificate presented for signing, the validation may pass. However, the browser's view of  this same certificate may be entirely different. As such, the CA signed a certificate that will be interpreted differently at the time of its use.
\end{enumerate}

Some of the interesting netsplit attack vectors together with publicly known examples are listed below. 
\section{Common Blockchain Bug Vulnerabilities} 

\subsection{Netsplit due to multiple client implementations}

If there are multiple blockchain client implementations running on the network, the question is whether different clients will
make different decisions about block validity. If, for the same and fixed blockchain state, a client written in Rust and a client
written in C++ deem a block as valid and invalid, respectively, the two types of nodes adopt a different view of the ledger (the end
result may be a temporary or a permanent split). See for instance the \href{https://www.ethernodes.org}{distribution} of different Ethereum 
blockchain client implementations running on the network. 

As such, the problem boils down to the problem of equivalence of software. On the parsing/language level, relevant is the realization that a 
protocol implementation is in fact an implementation of a {\it dialect} of that protocol, see \cite{darmouth-talk} 
 and the discussion on Postel's law ("Be liberal in what you accept and conservative in what you send"). When it comes to context/state-dependent
 processing, such as block validation in a blockchain, see how consensus rules are in fact unknown \cite{consensus-unknown}.
 As a consequence of such considerations, Bitcoin considers the reference implementation to be its specification, since the reference code ultimately best specifies the protocol \cite{devguide}. A specification, written say in a form of a PDF document, will likely not be complete, fully accurate or in sync: an actual implementation is in fact the source of truth.

Examples of this blockchain bug include a 2016 issue which showed consensus disagreement between \texttt{geth} and \texttt{Parity} Ethereum clients. In some 
edge cases, the state update function would behave incorrectly on one of the implementations and correctly on the other, i.e., reverting empty account deletions is not performed in case of an out-of-gas exception during the deletion. From the Nov 2016 Ethereum advisory \cite{nov2016eth}:

\begin{quote} Geth was failing to revert empty account deletions when the transaction causing the deletions of empty accounts ended with an an out-of-gas exception. An additional issue was found in Parity, where the Parity client incorrectly failed to revert empty account deletions in a more limited set of contexts involving out-of-gas calls to precompiled contracts\end{quote}

  Another example from 2013 is a failure of the \texttt{bitcoin-ruby} reimplementation client to meet the (buggy) Bitcoin reference implementation  behavior. For \texttt{SIGHASH\_SINGLE} (non-standard) transactions, if the number of inputs is greater than the number of transaction outputs, during signature verification, the transaction hash (over which the signature is computed) is wrongly calculated to be equal to \texttt{000...01}. While this is clearly incorrect behavior, the reimplementation clients failed to reimplement this bug, resulting in a fork risk. See the corresponding Bitcoin forum post \cite{bitcointalk-ruby}:

  \begin{quote}
    I forked webbtc.com on testnet yesterday, still not quite sure how I pulled that off, but I was playing with non-standard transactions. I also forked webbtc.com on mainnet today with transaction 315ac7d4c26d69668129cc352851d9389\\b4a6868f1509c6c8b66bead11e2619f

    It has two inputs, both using SIGHHASH\_SINGLE, and only one output. SignatureHash() in script.cpp has the following code:

if (nOut >= txTmp.vout.size())
\{
printf("ERROR: SignatureHash() : nOut=\%d out of range\\n", nOut);
return 1;
\}

The thing is, that's not actually an error. SignatureHash() returns the hash of the signature, and CheckSig() doesn't check the return code, so it acts as though the hash the signature signed was 00000000000000000000000000000000\\00000000000000000000000000000001 You can create a signature for that, and it works just fine and the transaction is considered valid.
  \end{quote}

If a single implementation behaves in an unexpected or unreasonable way for a certain edge case when processing a transaction or a block, such behavior is a candidate: the other implementation may perform a reasonable action in that same case, resulting in a difference in behavior. Manual comparison of the source code of multiple implementations, creating input tests to compare how the two implementations react and using cross-implementation fuzzing techniques are all good ways to identify bugs of this type.

\subsection{Netsplit due to execution environment discrepancies}
Even in the case where there is only one reference implementation of the blockchain client running on the network, it runs in 
different environments on different nodes. The environment specifics include architecture differences (e.g., 32-bit or 64-bit), 
operating system (Linux vs. Windows), locale settings (language, region), configuration, etc. Furthermore, the blockchain client may 
rely on the libraries that are installed on the system and automatically updated (without fixing the libraries' versions). The question 
is whether these differences can make the clients behave differently in terms of block validation rules. 

A good example of this are the two Bitcoin consensus issues by OpenSSL's ECDSA signature handling in 2014 and 2015. Recall that ASN.1 is a 
standard from 1984 for specifying abstract objects which then can be serialized. Such objects can then be converted to binary data using 
encoding of choice, such as BER (Basic Encoding Rules),  DER (Distinguished Encoding Rules), PER (Packed Encoding Rules), etc. As for BER,
given an object, there exist multiple BER encodings. To ensure uniqueness, one has to use DER, a subset of BER, for which there exist a 
unique serialization of any given object. 

OpenSSL's function to validate an ECDSA signature \texttt{ecdsa\_verify} used to accept BER-encoded ECDSA signatures. However,  this \href{https://github.com/openssl/openssl/commit/85cfc188c06bd046420ae70dd6e302f9efe022a9}{OpenSSL commit from late 2014} locks \texttt{ecdsa\_verify} down to
accepting only DER-encoded signatures. The commit is a reaction to \href{https://nvd.nist.gov/vuln/detail/CVE-2014-8275}{CVE-2014-8275}, 
which reported ineffective certificate blacklisting via recording fingerprints (if the BER encoding of the signature can be tweaked 
while remaining valid, the fingerprint can be changed as well and as such blacklisting can be bypassed).

In such a situation, if a subset of the nodes on the network upgrade the OpenSSL library and the remaining nodes do not, a netsplit
can occur. This would happen if a block with a BER-valid but DER-invalid signature would be mined. Upgraded nodes could not validate
this block, whereas the legacy nodes would not be prevented from validating it. See the explanation \cite{bitcoin-dev}:

\begin{quote}OpenSSL 1.0.0p / 1.0.1k was recently released and is being pushed out by various operating system maintainers.  My review determined that this update is incompatible with the Bitcoin system and could lead to consensus forks. [..] If you are running third-party or self-compiled Bitcoin Core or an alternative implementation using OpenSSL you **must not update OpenSSL**. [...] While for most applications it is generally acceptable to eagerly reject some signatures, Bitcoin is a consensus system where all participants must generally agree on the exact validity or invalidity of the input data.  In a sense, consistency is more important than "correctness".\end{quote}

The attack described above requires that nodes are running different OpenSSL versions. A good example of a netsplit vector 
that does not even require running different library versions was disclosed in July 2015 \cite{bitcoin-dev2}. Recall that
ASN.1 BER encoding is a rather complicated standard and it turns out that OpenSSL's BER parser will behave differently
depending on the platform (Windows/Linux/MacOS and 32-bit/64-bit). Specially crafted ECDSA signatures may or may not
validate depending on the platform the node is running on, resulting in blockchain forks. From the disclosure:

\begin{quote}One of the features of BER is the ability for internal structures to have a length descriptor whose size itself is up to 126 bytes (see X.690-0207 8.1.3.5). A 1 terabyte data structure would for example use a 5-byte length descriptor. However, there is no requirement to use the shortest possible descriptor, so even a 70-byte ECDSA signature could use a 5-byte length descriptor and be valid. Unfortunately, OpenSSL supports length descriptors only as long as their size is at most that of a C 'long int', a type whose size depends on the platform (Windows and 32-bit Linux/OSX have a 4-byte long int, 64-bit Linux/OSX have an 8-byte long int).\end{quote}

A blockchain client implementation should be as independent from the execution environment as possible. Reviewing for this type of bug includes  verifying whether the client fixes external software versions/commits, properly handling 32-bit and 64-bit discrepancies, any dependence on how operating system's system calls are implemented, whether locale differences play a role in consensus related code, etc. 

\subsection{Netsplit via block hash poisoning}

When a new block or a transaction is processed by a client, its hash is looked up in the client's store of already known block hashes. 
If the hash turns out to be known, the payload may be discarded, as it is assumed it was already processed. This is an optimization 
which prevents a single payload from being processed multiple times in a P2P network setting. 

It is important to note that the block or the transaction in question may have been invalid and rejected; regardless, its hash may be
recorded as already processed. As such, if a malicious participant manages to {\it invalidate} a legitimate block {\it without} modifying its hash, 
it becomes unclear what block the hash pertains to: the valid or the invalid one. To exploit this, a malicious participant without mining 
capabilities may listen on the network for newly mined blocks and attempt to quickly broadcast an invalidated variant of the same block.
The nodes that receive the invalidated block first will reject the actual truly valid block later on. The nodes that received the valid
block first will correctly accept the block, leading to a netsplit condition. 

Possible ways to invalidate a block without modifying its hash include:

\begin{itemize}
\item If the {\it overall hashing method} is vulnerable to a second pre-image attack. To clarify: it is assumed that the hash function (such as SHA-256) is resistant to second pre-image forgery, however, the overall hashing method includes data serialization, combining multi-field data and/or padding and these may render the construction less resistant to second pre-imaging. 
\item If there exists a block/transaction field, not included in the hash, but relevant for the validity decision. Such a parameter could be trivially modified to affect the block validity without changing its hash
\end{itemize}

As an example of the second-preimage problem on a block's hash, see Bitcoin's CVE-2012-2459. Consider how a block hash is derived. It is in fact a hash of the block header fields. The block header fields need to be combined together into a byte string and one of the fields is the Merkle root of the transaction set. When it comes to the transaction set, it is probable that the number of transactions will not correspond to a full binary tree leaf set (the number will likely be different than $2^k$ for some $k$). As such, the transaction set needs to be padded to complete the tree computation. The padding used in Bitcoin rendered this vulnerable to an attack \cite{forrestv}:

\begin{quote}
The Merkle hash implementation that Bitcoin uses to calculate the Merkle root in a block header is flawed in that one can easily construct multiple lists of hashes that map to the same Merkle root. For example, merkle\_hash([a, b, c]) and merkle\_hash([a, b, c, c]) yield the same result. This is because, at every iteration, the Merkle hash function pads its intermediate list of hashes with the last hash if the list is of odd length, in order to make it of even length.

And so, the Merkle root function can be effectively preimaged by changing the input so that one of the intermediate lists is of even length with the last two elements equal (where originally it was of odd length with a last element equal to the earlier mentioned two). As was later noted, this extends to any input length that is not a power of two: merkle\_hash([a, b, c, d, e, f]) == merkle\_hash([a, b, c, d, e, f, e, f]). Note that to maintain the same root hash, the only flexibility that exists is duplication of elements.

As a result, two blocks can easily be created that have the same block hash, though one can be valid and the other invalid, by duplicating one or more of the transactions in a way that maintains the Merkle root hash. Duplicating any transaction will make the block invalid, since the block double spends a certain past transaction output.
\end{quote}

Even though the underlying hash function is second preimage-resistant, it was still possible to second-preimage the construction. When validating an implementation for this type of bug, the question is whether it's possible to make a node consider a block invalid without changing the legitimate block's hash. Another issue that may make the hashing method vulnerable to collisions is the ambiguous serialization, see e.g. this old bug in AWS \cite{aws}. The question is whether there are collisions in operations that precede the actual hashing.

\subsection{Netsplit via unintended or pre-mature fork}

Suppose a blockchain client is scheduled for an upgrade and suppose there's a PR (pull request) that implements the change. 
When reviewing such a PR, there are two options when it comes to what it claims:

\begin{itemize}
\item[-] The PR claims to {\it not} change the rules on what constitutes a valid block
\item[-] The PR knowingly changes the block validity ruleset 
\end{itemize}

{\it If the PR is a non-forking change}, the important security question is whether the rules really remain 
unchanged, or not. Seemingly innocuous code updates, which do not appear to affect the block validation rules, 
may in fact affect and change the ruleset. A block that validates with, say, upgraded nodes and does not validate 
with the legacy nodes would cause a netsplit condition, resulting in a double-spend possibility on the blockchain 
users. 

In its most direct form, this bug is due to the fact that the upgrade simply (and unintentionally) changes
what a valid block is. As a good example, see this 2018 Bitcoin Cash vulnerability disclosure \cite{bitcoin-cash}:

\begin{quote}A critical vulnerability exists in bitcoin-abc that would allow anyone to bifurcate the Bitcoin Cash blockchain simply by sending an otherwise-harmless transaction. [...] With this change, the requirements for a signature's hash type were inadvertently loosened to allow the third-most significant bit to be set. Because this allows previously-invalid transactions to become valid, a valid block containing at least one of these transactions would result in a hard-fork.\end{quote}

A more subtle variant of this flaw is what looks like an innocuous underlying library update. The
Bitcoin 0.8 release, meant to be a non-forking change, which, among \href{https://github.com/bitcoin/bitcoin/blob/master/doc/release-notes/release-notes-0.8.0.md}{other things}, replaced BerkleyDB, the underlying database the client uses,  with LevelDB. This change
also removed a 10.000 Berkley DB lock limit, which imposed an (unknown) consensus rule: the number of transaction
inputs in a block was lower than expected. The switch to LevelDB removed this constraint and made previously invalid 
blocks valid. See \href{https://en.bitcoin.it/wiki/BIP_0050}{BIP 50}:

\begin{quote}
 A block that had a larger number of total transaction inputs than previously seen was mined and broadcasted. Bitcoin 0.8 nodes were able to handle this, but some pre-0.8 Bitcoin nodes rejected it, causing an unexpected fork of the blockchain. 
 [...]
 During this time there was at least one large double spend. However, it was done by someone experimenting to see if it was possible and was not intended to be malicious. 
 [...]
 With the insufficiently high BDB lock configuration, it implicitly had become a network consensus rule determining block validity (albeit an inconsistent and unsafe rule, since the lock usage could vary from node to node).
\end{quote}

To review for this type of issues, carefully look at all the changes the client is undergoing and assess whether they can change the consensus rules. If an underlying libraries are updated, inspect whether the changes in the library code imply changes in the client's behavior. 

{\it If the PR is a forking upgrade:} A network transition to a new ruleset can happen in several ways, options include:

\begin{itemize}
\item[-] Flagday cutover activation 
\item[-] Miner signaled activation
\end{itemize}

In Bitcoin these are called UASF (User Activated Soft Fork) and MASF (Miner Activated Soft Fork). In the flagday 
activation, clients are upgraded, but no change happens before a pre-set block height. It is assumed that a large
majority of the clients will upgrade before the activation. In the miner signaled activation, clients upgrade and
no change happens at all unless a sufficient number of miners signal readiness for moving to the new ruleset. 
Variants on these approaches are also possible, such as a combination thereof. To familiarize with these concepts, 
see \href{https://en.bitcoin.it/wiki/BIP_0034}{BIP 34} and \href{https://en.bitcoin.it/wiki/BIP_0009}{BIP 9} for MASF 
and \href{https://en.bitcoin.it/wiki/BIP_0008}{BIP 8} for the "combined" approach. See also this informative blog post \cite{lombrozo}
and various content on the  segwit adoption controversy. 

Regardless of the exact transitioning mechanism, there has to be an activation moment uniquely determined by all the 
upgraded clients on the network, where the old rules cease to be enforced and new rules take place. For a successful
fork, a large majority of the network should obey this unique moment and switch the rules. Any mistakes in this switch
result in {\it rule confusion} problems. The following two constraints need to hold:

\begin{itemize}
\item[] (1) \textbf{Pre-activation legacy rules consistency}: The client upgrade should not modify the legacy rules {\it before} the activation
\item[] (2) \textbf{Post-activation updated rules consistency}: New/updated rules should not leverage a legacy data structure or reuse legacy rule code {\it after} the activation
\end{itemize}

A violation of (1) would imply a netsplit {\it before} the agreed activation moment, when the disbalance between the upgraded 
and nodes that did not upgrade is stark, defying the purpose of the controlled rule transition. Violations of (2) results 
in incorrect and unintended consensus rules. 

Consider how a violation of (2) may happen and the same is true for (1). The client may reuse some of the legacy memory data structures,
which cause the inconsistency. Similarly, a legacy consensus-relevant function may be used inside upgraded rule validation.
See \cite{abc-reddit-writeup} on the Bitcoin Cash bug from 2019. The upgrade introduced a novel script operator, but relied on legacy code to determine if this is a signature validation operation 
(sigOp) or not:

\begin{quote} When OP\_CHECKDATASIG was introduced during the November upgrade, the procedure that counted the number of sigops needed to know if it should count OP\_CHECKDATASIG as a sigop or as nothing (since before November, it was not a signature checking operation). The way the procedure knows what to count is controlled by a "flag" that is passed along with the script. If the flag is included, OP\_CHECKDATASIG is counted as a sigop; without it, it is counted as nothing. Last November, every place that counted sigops included the flag EXCEPT the place where they were recorded in the mempool--instead, the flag was omitted and transactions using OP\_CHECKDATASIG were logged to the mempool as having no sigops.\end{quote}

To review for this type of issue, consider whether legacy data structures or functions are used in upgraded code and vice-versa, whether the newly introduced code and data structure changes affect the old rules in any way.

\subsection{Netsplit via branch confusion}

Even though at each moment only one chain of blocks is considered authoritative, PoW-based blockchain clients typically keep track of the full {\it block tree}. The block tree represents all possible ledger states the client knows about. When a new block is processed, if it connects to a non-authoritative branch of the block tree, the client needs to temporarily switch to that particular side branch, adopt that branches' transaction dataset and process the new block under those terms. 

Suppose, however, that branches can affect each other. This could happen as follows:

\begin{itemize}
\item \textbf{Cross-branch read}: Branch A mistakenly relies on data from branch B (incorrect computation on branch A).
\item \textbf{Cross-branch write}: Computation on branch B results in an unsolicited modification of branch A. 
\end{itemize}

The first issue could happen if the node is processing a block on a non-authoritative branch and the code mistakenly uses data from the authoritative branch. It may appear that branch confusion problems are simply incorrect computation on new blocks, however, they can be turned into netsplit attack vectors. This is because the (incorrect) computation happens assuming the node {\it sees the side branch}: if the node does not see the side branch, the computation remains unaffected. If a part of the nodes never sees the affecting side branch, that part of nodes will perform different computation than the remaining nodes. Cross-branch problems should be convertible to netsplit attack vectors. 

Consider two examples of this interesting and somewhat hard to find bug. The first one is the BLOCKHASH issue \cite{blockhash} in Ethereum from 2015. The BLOCKHASH(n) function returns hashes of blocks' predecessors. When called on a side branch that is non-authoritative, the function mistakenly returned the authoritative branches' previous block hashes:

\begin{quote} Both C++ (eth) and Go (geth) clients have an erroneous implementation of an edge case in the Ethereum virtual machine, specifically which chain the BLOCKHASH instruction uses for retrieving a block hash. This edge case is very unlikely to happen on a live network as it would only be triggered in certain types of chain reorganisations (a contract executing BLOCKHASH(N - 1) where N is the head of a non-canonical subchain that is not-yet reorganised to become the canonical (best/longest) chain but will be after the block is processed).\end{quote}

Consider also Bitcoin's Transaction Overwriting CVE-2012-1909. The issue follows from the fact that the miner could duplicate a coinbase transaction, see \href{https://github.com/bitcoin/bips/blob/master/bip-0030.mediawiki#motivation}{BIP 30}:

\begin{quote}So far, the Bitcoin reference implementation always assumed duplicate transactions (transactions with the same identifier) didn't exist. This is not true; in particular coinbases are easy to duplicate, and by building on duplicate coinbases, duplicate normal transactions are possible as well.\end{quote}

See also \cite{stackexchange-tx-overwriting} for some more details on netsplit aspect of this issue are explained: 

\begin{quote}The problem was that in case a duplicate transaction was created in a side branch that is afterwards reverted, and is only seen by a certain portion of the network, a fork risk exists. The nodes A that have seen the duplicate transactions and its reversal, will consider the original transaction unspendable (as it was overwritten and subsequently removed from their transaction database in the reorganisation), while nodes B that did not see the duplicate would consider the original spendable. When the original transaction is spent afterwards, and a majority of the network is in B, the network will split, as the A nodes will consider the chain created by B as invalid.\end{quote}

While working on the side branch, the client affected the future authoritative branch, by deleting one of the transactions in the past.

When reviewing for this type of issue, investigate whether block and transaction processing functions reference any cross-branch memory. The authoritative branch memory may be referenced instead of the temporary branch when a new block is added on the temporary branch. Consider whether temporary blockchain data memory caches are correctly substituted during jumping between branches. Finally, consider whether the global client's view of the blockchain can be affected during reorgs.

\subsection{Improper timestamp validation}

Recall the reason that block headers contain timestamps: To reward miners for their work, each block generates new coins and the miner who generated the block has control over the generated coins. Given that each block generates new coins, there needs to exist a mechanism that controls the rate at which blocks are mined. This can be a target number of blocks per a time period. To measure how many blocks were mined in a time interval, the networks needs to attach a timestamp to each block.

Every n blocks, clients perform an evaluation of how much time it took to mine those n blocks. Based on this evaluation, clients adjust the expected mining difficulty, used as a criterion for block validity. Mining difficulty can be expressed as a value the PoW hash needs to be less than. If it took longer than a pre-set interval to generate those n blocks, the difficulty decreases and if it took shorter, the difficulty increases. Owing to timestamp and difficulty parameters inside the block header fields, the network can self-control the rate at which new blocks are mined.

To verify for issues in time stamp validation, a good starting point is to look for deviations from what \href{https://github.com/bitcoin/bitcoin/blob/ab4e6ad7/src/validation.cpp#L3439}{Bitcoin} (and \href{https://github.com/ethereum/go-ethereum/blob/c4b7fdd2/consensus/ethash/consensus.go#L25)}{Ethereum} do. As for Bitcoin:

\begin{itemize}
\item[-] The timestamp has to be strictly greater than the timestamp median of the previous 11 blocks
\item[-] It also has to be less than or equal to the network adjusted time + 2 hours
\end{itemize}

When it comes to basic timestamp validation problems, a simple omission of either a lower or upper bound opens the network to severe attacks. In that case, mining a block with a degenerate time stamp (e.g., equal to the minimal or maximal possible value) will likely prevent any other blocks from being added. For instance, difficulty may be manipulated to an unrealistically high bar, or the subsequent time stamps will be expected to be higher than the maximal possible timestamp value. 

If both upper and lower bound are enforced, but the lower bound does not depend on previous blocks' timestamps, a netsplit attack can arise. While possibly  difficult to pull off in pratice, if a miner broadcasts a block on the brink of expiry, some nodes may accept it and other nodes may not, depending on the node's view of local time, resulting in a chain discrepancy. Note that the same attack does not work for the upper bound, as the block that is on the brink of becoming valid (with timestamp close to the upper bound) will become valid and the network will get a chance to auto-correct during the timestamp validity interval (unless the initial rejection blacklisted the block). 

Another attack related to timestamps is known as the Time Warp attack \cite{timewarp1}.  As mentioned above, the timestamps are necessary to determine mining difficulty.  For instance, in Bitcoin, every 2016 blocks, the mining nodes adjust the mining difficulty.  The difficulty is calculated by counting the time difference between the two blocks at the beginning and the end of the mining interval. The idea of the attack is  to control the timestamps at the ends of this interval and manipulate the difficulty.  For instance, as for the end of the interval, the attacker would use a maximal possible timestamp, so the block is still accepted. This extends the time used in the difficulty calculation so that the nodes are tricked in to thinking it took longer to reach 2016 blocks than it really did.  The final goal is to lower the difficulty. This could be done on the main chain but also on a fork, with a goal of having the forked chain promoted as the authoritative chain. 

This attack is called the Time Warp attack because, the very next block added after the forged timestamp block would likely revert back in time.  This is acceptable because of rule 1 above in which the timestamp needs to only be greater than the median of the previous 11 blocks. See the discussion on longest vs. most-work chain \cite{longest1, longest2}.

Finally, an attacker with a significant percentage of the hashrate, such as colluding miners or a very powerful attacker, can try to mine blocks very quickly by spacing blocks out by the minimum, 1 second or so. This is possible, as the timestamp needs to be greater than the median of the previous block's timestamps. However, the final block in the sequence interval (e.g., 2016th block) would be pushed as far in the future as possible. The end result is that the time between the starting and ending block is high, resulting in a mining difficulty drop.  This way, at least in theory, miners could very rapidly mine coins for themselves, resulting in a significant inflation in the supply of coins available.  This attack also requires a significant portion of the hashrate which increases the difficulty of performing this attack and is unlikely to go unnoticed. 

Relevant here are also the Time Jacking attack \cite{timejacking}, although it is somewhat Bitcoin specific since it depends the ability of peers to affect other peers' time. It includes tweaking other nodes' time, for the attacker's advantage, and this shouldn't be possible for many other implementations. See the Time Warp attack on the Verge \cite{theverge} as an example of a successful time-based attack in practice. 

\subsection{Integer underflow/overflow}

The Bitcoin's value overflow incident \cite{overflow} refers to an event where an integer overflow in a transaction allowed two transaction outputs to be created. Each output received 9223372036854277039 Satoshis, or ~92.2 billion coins.  As a point of reference, the max value of a 64 bit signed integer is 9,223,372,036,854,775,807. The data type used to store the amount is a C \texttt{int64} type, a signed 64 bit integer. Converted to an integer, the numbers in the transaction outputs will overflow an int64 data type when they are summed.

Prior to this incident, checks were not performed to prevent overflowing the summed integer of all the transaction output values, see the original code points,  \href{https://sourceforge.net/p/bitcoin/code/131/tree/trunk/main.cpp#l1011}{main.cpp} and \href{https://sourceforge.net/p/bitcoin/code/131/tree/trunk/main.h#l474}{main.h}.  A check was performed to  prevent output values from being negative only but not the sum of the values, see the \href{https://sourceforge.net/p/bitcoin/code/131/tree/trunk/main.h#l518}{GetValueOut} function. A \href{https://sourceforge.net/p/bitcoin/code/131/tree/trunk/main.cpp#l1071}{check} was also performed that total output was not greater than total value.  This allowed a situation where individual outputs were very large, but their sum was not greater than the total input value, so the transaction passed the checks.

After this incident, checks on both the input value and output values are performed.  A new \texttt{MAX\_MONEY} constant was defined that limits the maximum size of input and output values.  The sum of all inputs and the sum of all outputs are also checked to be less than this value as well.  Additionally, no input or output value can be less than 0, see the related \href{https://github.com/bitcoin/bitcoin/commit/d4c6b90ca3f9b47adb1b2724a0c3514f80635c84#diff-118fcbaaba162ba17933c7893247df3aR1013}{diff}. 

It is important to prevent integer overflows when calculating input and output amounts dealing with values in transactions.  In this case, an overflow passed the original test in place but it allowed very large amounts of bitcoin to be created as a result.  The strategy of limiting the maximum size of a transaction, checking that values after summation are not negative, and additionally, checking that the total sum of inputs and outputs are separately checked to not pass a maximum value is an effective way to solve this issue by asserting that transactions can all be represented in a certain size integer by using only a subset of the representable numbers.  That way, no two values from the usable numbers can overflow the integer.

\subsection{Merkle tree implementation issues}
The Bitcoin paper introduces a concept of SPV (Simplified Payment Verification) nodes and this concept is reused in one form or another in many blockchain implementations \cite{SPV}. The goal is to allow nodes to verify transactions without downloading full blocks. This is achieved by relying on the concept of Merkle trees. A Merkle tree is nothing but an efficient mechanism to efficiently/succinctly prove set membership. To validate if a given transaction is included in a block, a node does not have to download the full block. Rather, the node can just verify the short Merkle proof and become convinced that the transaction is indeed included in the block. 

For a detailed description on how a Merkle tree is generated in Bitcoin, refer to \href{https://bitcoin.org/en/blockchain-guide#transaction-data}{this} page. In general, transactions' TXIDs (Transaction IDs) are hashed and concatenated in a tree-like structure, resulting in top tree node, the Merkle root. The Merkle root of a block's transaction set is a unique hash determining the transactions and is included in the header of every block. To produce a short proof without needing the whole block, a transaction can be verified by taking its TXID and sibling TXIDs required to reconstruct the Merkle root (around $log_2 N$ nodes, where $N$ is the number of transactions in the set).  The verifier can then verify that a transaction is inside of a block. 

When implemented naively, Merkle trees introduce security problems in blockchain. Common issues include:

\begin{itemize}
\item[-] Leaf-node weaknesses, or, lack of differentiation between leaf and internal Merkle tree nodes
\begin{itemize}
\item[-] Internal Merkle Tree nodes interpreted as transactions.  This could allow internal nodes of a block's Merkle tree to be interpreted as transactions and relayed.  This was the case with BTC Relay \cite{btcrelay}.  Note that a random node is unlikely to be a valid transaction, however.  The proposed solution to this issue was to ignore transactions that happen to be exactly 64 bytes, or the same length as two concatenated 32 byte hashes, to resolve the issue.
\item[-] The reciprocal issue, transactions interpreted as valid internal nodes of a block's Merkle Tree.  This may allow attackers to construct a valid proof of a fake transaction and use this to trick a victim relying on SPV proof of validity.  This is a complete violation of the assertion that a Merkle Tree proof is meant to provide which shows that a transaction is valid and exists inside of a some block (details of this issue are below).
\end{itemize}
\item[-] Merkle transaction set padding issues, see the bug in the block hash poisoning section of this paper. 
\end{itemize}

Other issues in this context (although not necessarily inherently related to Merkle trees) include a Monero bug from 2014, see \cite{monero1} \cite{monero2}.

\textbf{Leaf-node weakness}. Below we complement the previous description   of this attack with additional clarification \cite{sdlmerkle}. 

In general, the actual size of the Merkle Tree is implicit.  Normally, it would be $log_2(N)$, where N is the number of transactions in the block.  So, the amount of data to provide for such a proof can be much smaller than the block size.  However, the actual depth of the tree is implicit in the amount of data provided as a proof.  As a result of this, it is possible to submit a proof that treats a specially crafted transaction as an inner node of the tree.  Inner nodes are 32 bytes.  An attacker can submit a transaction that is exactly 64 bytes that can be reinterpreted as a pair of \texttt{TXID} values used in an inner node.  The second half of the specially crafted, 64 byte, transaction is also a valid \texttt{TXID} for a fake transaction.  The fake transaction can be a payment to a victim.  This proof tricks the victim to verify that they received a fake transaction.  The attack is performed in two stages.

The cost of this attack computationally comes in two phases.  In the first phase the attacker must brute force 72 bits and in the second the attacker must brute force 40 bits.  The attacker should have at least $2^{36}-1$ Satoshi.

{\noindent{\textbf{Phase 1:}}}

In this phase at least 72 bits must be brute forced.  Note that this does not require brute forcing all 32 bytes of the \texttt{TXID}.We will brute force the \texttt{TXID} of a fake payment, {\it F}.  Our specially crafted transaction will be $T$.  The first half of $T$ will be $T_1$ and the second half $T_2$.

We begin by constructing our fake transaction $F$.  We will then attempt to brute force its \texttt{TXID} and let $T_2$be equal to it.  Below is a table describing the bits of $T_2$. As for $T$, first phase:

\begin{center}
 \begin{tabular}{|| c c c ||} 
 \hline
 Field & Byte Offs. & Brute F. Bits \\
 \hline
 T1 & 0-32 & 0 \\
 TXID (second)	&	 32-37	&	 0			\\
 VOUT		&	 37-41	&	 17 of 32		\\
 ScriptSigSize	&	 41	&	 8			\\
 Empty Script	&	 42	&	 0			\\
 Sequence	&	 42-46	&	 0			\\
 Output Count	&	 46	&	 8			\\
 Value		&	 47-55	&	 29 of 64		\\
 ScriptPubKey Size &	 55	&	 8			\\
 ScriptPubKey	&	 56	&	 0			\\
 lock\_time	&	 60-64	&	 2			\\
\hline
\end{tabular}
\end{center}

In the second half of $T$, some bits can be free which reduces the total number of bits that must be brute forced.
Importantly, the following fields shown above must be brute forced.

\begin{itemize}
\item[-] 2 bits of \texttt{lock\_time} such that the value in lock time is less than $2^{30}$. This implies the \texttt{lock\_time} has passed.
\item[-] 8 bits of \texttt{ScriptPubKeySize} (equal to 0).
\item[-] 29 bits of \texttt{Value}.  The attacker should have at least $2^{36}-1$ satoshi so the remaining bits should be 0. The other 35 bits can be any value.
\item[-] 8 bits of \texttt{Output Count}.  (equal to 0).
\item[-] 8 bits of \texttt{ScriptSigSize}. (equal to 0).
\item[-] 17 bits of \texttt{VOUT}.
\end{itemize}
To perform this phase of the attack, after constructing the fake transaction $F$ we can iterate through \texttt{lock\_time} and the input script until a suitable $T_2$ is found.

{\noindent{\textbf{Phase 2:}}}

The first half of the transaction $T_1$ must now be completed.  The input hash value for our transaction $T$ lies on both $T_1$ and $T_2$.  The last 40 bits of the input \texttt{TXID} lie in $T_2$.  These 40 bits have been previously selected in phase 1 when $T_2$ was brute forced.  In this phase, we must complete a valid transaction using both halves such that the input \texttt{TXID} contains the last 40 bits selected in the prior phase.

Additionally, the \texttt{VOUT} value was chosen previously and this will also be used in this phase.  Below is a table showing what is brute forced in this state.


\begin{center}
 \begin{tabular}{|| c c c ||} 
 \hline
 Field & Byte Offs. & Brute F. Bits \\
 \hline
 Version		& 0-4		& 0			 \\
 Input Count		& 4		& 0			\\
 TXID			& 5-32		& 0			\\
 TXID			& 32-37		& 40			\\
\hline\hline
 rest of T2		& 37-64		& 0			\\
\hline
\end{tabular}
\end{center}

As you can see, the final 40 bits of \texttt{TXID} were placed $T_2$.  We must create a transaction, $P$, with a matching transaction ID value, a matching $Q$ index, and for the amount $A$.

To perform this step, we must create another transaction, $P$ with the final 40 bits of its \texttt{TXID} matching our present value 
in $T_2$.  We will therefore create a transaction for value $A$ and with index $Q-1$ that matches.  We can iterate 
through \texttt{lock\_time} to try to find the matching value.  The total number of bits to brute force in this stage of the 
attack is $2^{40}$. 

The chain of transactions will finally be: \texttt{P -> T -> E} where E is a separate transaction that will capture the
output from $T$.

To conclude, the Merkle Trees used to perform quick verification of a transaction present the risk of transactions being interpreted as
inner nodes which potentially allows a valid proof of a fake transaction.  Some ways to fix this include:

\begin{itemize}
\item[-] Check if inner nodes of a Merkle Tree are valid transactions.  If they are, flag presence of 64 byte transaction.
\item[-] Use a Merkle proof for both a transaction to be asserted and the coinbase transaction. The trees should be the same height. 
\item[-] Use inclusion proof on final transaction in a block in order to be able to compute the Merkle tree height.
\item[-] Include Merkle Tree height inside of a block.
\end{itemize}

\subsection{Storage exhaustion in block or transaction processing}

This is a somewhat straightforward Denial of Service attack vector. Blockchain nodes keep track of known blocks 
and tranasctions. The store that keeps uncofirmed transactions is a mempool. If there is no limit to how much a 
certain data store can keep, the node may run out memory or storage and stop functioning. For the attack to be 
feasible, it is necessary that pushing data onto nodes can be done for free (or at a manageable cost) from an 
attacker's point of view. 

Consider for instance blocks: the cost of generating valid blocks includes solving the proof of work puzzle at some difficulty level 
and, as such, the cost of adding new blocks to a node should be high. Nonetheless, overloading block storage was a DoS concern in the past:

\begin{itemize}
\item[-] \textbf{DoS via excessive orphan blocks}: Each block header contains a field that specifies its parent block hash. If a node receives  a new block but does not know of its parent block, it is difficult for the node to verify whether the block's proof of work  is correct (as it is unclear where in the blockchain the block belongs and what difficulty level is required  for the PoW proof).  Such blocks are also called {\it orphan blocks}. Nodes keep orphan blocks in hope that they will connect to the chain once their parent blocks become known. The early Bitcoin client (before 0.9.0) did not impose any limit on the orphan block store \cite{patvarilly}. A limit was implemented in this \href{https://github.com/bitcoin/bitcoin/commit/bbde1e99c89392}{commit}  in early 2014. 
\item[-] \textbf{DoS via excessive blocks branching off of an early blockchain point}: Another idea to bypass the proof of work difficulty is to mine many blocks which fork from (very) early blocks. Blocks connecting to early points in the blockchain need to satisfy minimal difficulty and are cheap to generate. This attack is mitigated by inclusion of \href{https://github.com/bitcoin/bitcoin/commit/d8b4b49667}{checkpoints} which reject blocks at very old heights.
\end{itemize} 

Similarly, as for transaction stores:

\begin{itemize}
\item[-] \textbf{Transaction mempool issues}: During the summer of 2015, a spam campaign against Bitcoin was performed \cite{spam}. The campaign was advertised as a "stress test" and it attempted to overload different aspects of the Bitcoin network, see also \cite{spam-study}. One of the results was a \href{https://www.reddit.com/r/Bitcoin/comments/3ny3tw/with_a_1gb_mempool_1000_nodes_are_now_down/}{significant pressure} on nodes' transaction mempools, which DoS-ed a large number of nodes on the network. See the \href{https://bitcoin.org/en/release/v0.12.0#memory-pool-limiting}{release notes} for Bitcoin v12.0.0, which resolved this by introducing a transaction ejection strategy. 
\item[-] \textbf{DoS via excessive orphan transactions}: Similarly to orphan blocks, orphan transactions are transactions which reference an unknown parent transaction. \href{https://en.bitcoin.it/wiki/CVE-2012-3789}{CVE-2012-3789} discusses a lack of limits on the number (and size) of orphan transactions a node can store. Peers may be flooded by orphan transactions until a node's memory is filled. 
\end{itemize}

It is important to enumerate {\it all} data stores the clients keep when reviewing for this type of issues
and to carefully assess the cost of adding data to nodes in all of the edge cases.

\subsection{CPU exhaustion in block or transaction processing}

While any endpoint exposed on the network is open to trivial packet flood DDoS attacks, a non-trivial CPU 
exhaustion attack achieves a similar goal with little network traffic. The target node is forced to perform 
a large number of operations per comparatively little network data originating from the attacker. The ratio 
between the amount of data the attacker needs to send and the number of operations the server runs determines
the effectiveness of the DoS vector. 

As such, the problem boils down to counting the number of operations executed by the node per the amount of 
data the attacker sends, in all possible scenarios. Transaction and block validation is complex, depends on 
the state of the ledger and includes performing cryptographic operations. This makes it a fertile ground for CPU 
exhaustion issues. There are two common vectors: 

\begin{itemize}
\item[-] \textbf{Algorithmic complexity attacks}: The question is whether any of the exposed processing/ingestion algorithms 
execute in, say $n^2$ time, where $n$ is the number of inputs the attacker supplied. This could be a 
corner/worst case of the algorithm execution, similar to how Quicksort's worst case is $n^2$
and the average case is the acceptable $n\cdot log n$ of operations, or how hash tables degrade on collisions,
see \cite{dospaper} for many examples. In the case of blockchain, the question becomes: what is the *worst-case complexity* for *all the algorithms* exposed by the blockchain client ingestion endpoints? 
\item[-] \textbf{Excessive number of cryptographic validations}: Cryptographic operations such as signature validations are CPU-intensive 
and nodes should restrict the number of operations they will execute. See Bitcoin's \href{https://en.bitcoin.it/wiki/Common_Vulnerabilities_and_Exposures#CVE-2010-5138}{CVE-2010-5138} and the discussion on the Bitcointalk forum in this regard \cite{sdl1}. 
\end{itemize}

A interesting example of an algorithmic complexity attack was identified in early Bitcoin's handling of orphan transactions, see the second part of \href{https://en.bitcoin.it/wiki/CVE-2012-3789}{CVE-2012-3789}. Consider the inherent complexity of dealing with orphan transactions, that is, transactions whose parent transactions are unknown. Keeping orphan transactions requires verifying if newly received (non-orphan) transactions make any of the known orphans non-orphans. Now, each transaction that needs to be un-orphaned may in turn make other orphans become non-orphans and this process needs to be applied recursively. 

To make the process efficient, Bitcoin relies on a C++ map \href{https://github.com/bitcoin/bitcoin/blob/v0.6.0/src/main.cpp#L2541}{\texttt{mapOrphanTransactionsByPrev}} that links parent transaction hashes with their respective (child) orphan transactions. A newly received (non-orphan) transaction can be efficiently looked up in such a map and potentially un-orphan orphans that depend on it. Now, when an orphan is deleted from the orphan store, the \texttt{mapOrphanTransactionsByPrev} mapping needs to be pruned as well. Each parent-orphan entry pointing to that orphan needs to be deleted. 

\href{https://en.bitcoin.it/wiki/CVE-2012-3789}{CVE-2012-3789} makes the victim node store a degenerate form of the \texttt{mapOrphanTransactionsByPrev} mapping and then triggers a deletion of a chosen orphan. Since the mapping is degenerate, the number of operations to remove the traces of the orphan inside the map becomes huge. See \cite{sdl2blog} for more details on how this works and how the degenerate map looks like. 

Other algorithmic complexity issues in Bitcoin include quadratic complexity run times in Bitcoin scripting language and transactions that take five hours to verify \cite{sdl3}, \cite{sdl4}. 

\subsection{Other blockchain client implementation level issues}

{\textbf{P2P network bandwidth exhaustion/spam}: It should not be possible for nodes inside a P2P network to get the network to relay meaningless bloat data, as this would impact the legitimate block and transaction propagation speeds. Even more so, meaningless data should not be stored, to avoid network-wide storage exhaustion problems.

An interesting bug likely to plague other blockchains is Bitcoin's \href{https://www.cvedetails.com/cve/CVE-2013-4627/}{CVE-2013-4627}.  First, the attacker takes a legitimate serialized message (e.g., a transaction message) that can be transferred over the network, parsed and accepted by participants.  Next, the bloat data is carefully added to the legitimate serialized message, so that the message {\it does not get invalidated}. How can this be done? Consider the fact that widely used serialization methods such as msgpack or protobufs support ignoring unknown fields. The ability to ignore unknown fields is a legitimate design requirement: an old parser should be able to process serialized messages generated by newly released code, by ignoring the unknown fields. If the library is configured to ignore unknown fields, the serialized message with an unknown field containing bloat will stay valid and be relayed on the network. 

Special-purpose or custom serialization code will also often tolerate junk data, one way or another. Bitcoin relies on custom serialization code, inserted to classes meant to be (de)serialized via C macros. During deserialization of composite objects, deserialization functions of lower level objects get called recursively (reminiscent of the Composite OOP design pattern). As for the code vulnerable to CVE-2013-4627, a deserialization code path for a newly received transaction can be observed \href{https://github.com/bitcoin/bitcoin/blob/09e437ba/src/main.cpp#L3506}{here}, calling the \href{https://github.com/bitcoin/bitcoin/blob/09e437ba/src/serialize.h#L1093}{overloaded} \texttt{CDataStream}'s  >> operator only to land in \texttt{CTransaction}'s \texttt{Unserialize} method, defined by the \href{https://github.com/bitcoin/bitcoin/blob/09e437ba/src/main.h#L481}{\texttt{IMPLEMENT\_SERIALIZE}} macro inside the \texttt{CTransaction} class:

\begin{verbatim}
   IMPLEMENT_SERIALIZE
    (
        READWRITE(this->nVersion);
        nVersion = this->nVersion;
        READWRITE(vin);
        READWRITE(vout);
        READWRITE(nLockTime);
    )
\end{verbatim}

The \texttt{READWRITE} macros are used during deserialization: once the \texttt{nLockTime} is read out from the data stream, the deserialization ends and ignores any other data that comes after. As such, it is possible to add an arbitrary junk data suffix to a valid serialized transaction. In Bitcoin, the serialized message was also saved to disk. An actual attack on the Bitcoin network due to this bug was discussed on the Bitcointalk forum \cite{serialization}. Mitigations for this issue include (1) not tolerating junk data by the deserialization code (2) the message size should be bounded and (3) messages should be first deserialized and then serialized again before relaying, to avoid relaying any eventual bloat. 

\section{Conclusion}

As any particular technology matures, the set of common security problems specific to that technology stabilizes and
can be subject to classification. This was the case for many other technologies in the past, such as web applications
or network setups. At this stage, web application developers and security bug hunters know what types of security
issues are to be expected. While blockchain in its variants is still a novel technology, the first decade of Bitcoin
provides a peek into what common bug templates look like.

In this paper\footnote{This work would not have been possible without: \textbf{Jeff Dileo, Jennifer Fernick, Mason Hemmel, Ava Howell, Ephrayim Kishko, Eric Schorn, Javed Samuel, Thomas Pornin and David Wong}}, we set out to help an aspiring blockchain security bug hunter by providing ten bug classes to look for
in their target blockchain client and also blockchain developers to know what type of bugs to be on the lookout for.


\begin{thebibliography}{100}

\bibitem{DASP} Decentralized Application Security Project (DASP), David Wong, Mason Hemmel \url{https://dasp.co/}

\bibitem{smartcontract} Nicola Atzei, Massimo Bartoletti, and Tiziana Cimoli,  A survey of attacks on Ethereum smart contracts

\bibitem{trailofbits} Trail of Bits, 246 Findings From Our Smart Audits Contract, An Executive Summary, \href{https://blog.trailofbits.com/2019/08/08/246-findings-from-our-smart-audits-contract-an-executive-summary}{URL}

\bibitem{ptacek}Thomas H. Ptacek, Timothy N. Newsham, Insertion, Evasion and Denial of Service: Eluding Network Intrusion Detection, 1998


\bibitem{bitcoincve}Bitcoin Common Vulnerabilities and Exposures, Bitcoin archives,  \href{https://en.bitcoin.it/wiki/Common_Vulnerabilities_and_Exposures}{URL}

\bibitem{bgp} Earl Zmijewski, Reckless Driving on the Internet, Blog post \href{https://dyn.com/blog/the-flap-heard-around-the-world/}{URL}

\bibitem{x509} PKI Layer Cake: New Collision Attacks Againstthe Global X.509 Infrastructure, Dan Kaminsky, Meredith L. Patterson and Len Sassaman

\bibitem{masteringbitcoin}  Andreas M. Antonopoulos,  Mastering Bitcoin

\bibitem{darmouth-talk} Len Sassaman and Meredith L. Patterson, Towards a formal theory of computer insecurity: a language-theoretic approach, Darmouth College  \href{https://www.youtube.com/watch?v=AqZNebWoqnc}{talk}

\bibitem{consensus-unknown} Peter Wuille, Why is it so hard for alt clients to implement Bitcoin Core consensus rules? \href{https://bitcoin.stackexchange.com/questions/54878/why-is-it-so-hard-for-alt-clients-to-implement-bitcoin-core-consensus-rules}{URL}

\bibitem{devguide} Bitcoin protocol  \href{https://en.bitcoin.it/wiki/Protocol_documentation}{documentation}

\bibitem{nov2016eth} Security alert [11/24/2016]:  \href{https://blog.ethereum.org/2016/11/25/security-alert-11242016-consensus-bug-geth-v1-4-19-v1-5-2/}{Consensus bug in geth v1.4.19 and v1.5.2}

\bibitem{bitcoin-dev} Gregory Maxwell, OpenSSL 1.0.0p / 1.0.1k incompatible, causes blockchain rejection  \href{https://lists.linuxfoundation.org/pipermail/bitcoin-dev/2015-January/007097.html}{URL}

\bibitem{bitcoin-dev2} Peter Wuille, Disclosure: consensus bug indirectly solved by BIP66  \href{https://lists.linuxfoundation.org/pipermail/bitcoin-dev/2015-July/009697.html}{URL}

\bibitem{bitcointalk-ruby} Peter Todd, A cautionary note: I just forked webbtc.com/bitcoin-ruby via two different ways \href{https://bitcointalk.org/index.php?topic=260595.0}{URL}

\bibitem{forrestv} Forrestv, [Full Disclosure] {CVE-2012-2459 Block Merkle calculation exploit}  \href{https://bitcointalk.org/index.php?topic=102395.0}{URL}

\bibitem{aws} Colin Percival, {AWS Signature Version 1 is Insecure}  \href{https://www.daemonology.net/blog/2008-12-18-AWS-signature-version-1-is-insecure.html}{URL}

\bibitem{bitcoin-cash} Theuni, {Bitcoin cash disclosure} \href{https://github.com/mit-dci/cash-disclosure/blob/master/bitcoin-cash-disclosure-04252018.txt}{URL}

\bibitem{lombrozo} Eric Lombrozo,  {Forks, Signaling and Activation}   \href{https://medium.com/@elombrozo/forks-signaling-and-activation-d60b6abda49a}{URL}

\bibitem{abc-reddit-writeup}  FerriestaPatronum, {ABC Bug Explained} \href{https://old.reddit.com/r/btc/comments/bp1xj3/abc_bug_explained/}{URL}

\bibitem{blockhash} Gustav Simonsson, {Security alert: Implementation of BLOCKHASH instruction in C++ and Go clients can potentially cause consensus issue}  \href{https://blog.ethereum.org/2015/10/22/security-alert-implementation-of-blockhash-instruction-in-c-and-go-clients-can-potentially-cause-consensus-issue-fixed-please-update/}{URL}

\bibitem{stackexchange-tx-overwriting} {More on BIP30 background}  \href{https://bitcoin.stackexchange.com/questions/5903/where-can-i-learn-more-about-bip30-namely-the-exploit-and-the-background-discus/5905}{URL}

\bibitem{timewarp1} ArtForz, {Possible way to make a very profitable 50 plus ish attack for pools?}  \href{https://bitcointalk.org/index.php?topic=43692.msg521772\#msg521772}{URL}

\bibitem{longest1} David A. Harding, {Bitcoin paper errata and details} \href{https://gist.github.com/harding/dabea3d83c695e6b937bf090eddf2bb3}{URL}

\bibitem{longest2} Ava Howell, {The Longest Blockchain is not the Strongest Blockchain}  \href{https://cryptoservices.github.io/blockchain/consensus/2019/05/21/bitcoin-length-weight-confusion.html}{URL}

\bibitem{timejacking} Alex Boverman, {Timejacking and Bitcoin}  \href{http://culubas.blogspot.com/2011/05/timejacking-bitcoin_802.html}{URL}

\bibitem{theverge} Daniel Goldman, {The Verge Attack, explained}  \href{https://blog.theabacus.io/the-verge-hack-explained-7942f63a3017}{URL}

\bibitem{overflow} {Bitcoin Overflow Incident}  \href{https://en.bitcoin.it/wiki/Value\_overflow\_incident}{URL}

\bibitem{monero1} Jan Macheta, Sarang Noether, Surae Noether and Javier Smooth,  Counterfeiting via Merkle Tree Exploits withinVirtual Currencies Employing the CryptoNoteProtocol

\bibitem{monero2} ferretinjapan, {What happened at block 202612}  \href{https://monero.stackexchange.com/questions/421/what-happened-at-block-202612}{URL}


\bibitem{btcrelay} Andrew Miller, {Security Audit of BTC Relayimplementation},  \href{http://soc1024.ece.illinois.edu/BTCRelayAudit.pdf}{URL}

\bibitem{patvarilly} Pat Varilly, {Denial of Service using orphan blocks?}  \href{https://bitcointalk.org/index.php?topic=23266.0}{URL}

\bibitem{spam} {July 2015 Flood Attack}  \href{https://en.bitcoin.it/wiki/July\_2015\_flood\_attack}{URL}

\bibitem{sdlmerkle} Sergio Demian Lerner, {Leaf Node Weakness in Bitcoin Merkle Treee design}  \href{https://bitslog.com/2018/06/09/leaf-node-weakness-in-bitcoin-merkle-tree-design/}{URL}
  
\bibitem{spam-study} Khaled Baqer, Danny Yuxing Huang, Damon McCoy,  Nicholas Weaver, Stressing Out: Bitcoin Stress Testing

\bibitem{dospaper} Scott A. Crosby,  Dan S. Wallach, Denial of Service via Algorithmic Complexity Attacks 

\bibitem{sdl1} Sergio Demian Lerner, {New Bitcoin vulnerability: A transaction that takes at least 3 minutes to verify} \href{https://bitcointalk.org/?topic=140078}{URL}

\bibitem{sdl2blog} Aleksandar Kircanski, {Bitcoin Orphan Transactions and CVE-2012-3789}  \href{https://cryptoservices.github.io/fde/2018/12/14/bitcoin-orphan-TX-CVE.html}{URL}

\bibitem{sdl3} Sergio Demian Lerner, {New Quadratic Delays in Bitcoin Scripts}  \href{https://bitslog.wordpress.com/2017/04/17/new-quadratic-delays-in-bitcoin-scripts/\#comment-13027}{URL}

\bibitem{sdl4} Sergio Demain Lerner, {A Bitcoin transaction that takes 5 hours to verify}  \href{https://bitslog.wordpress.com/2017/01/08/a-bitcoin-transaction-that-takes-5-hours-to-verify/}{URL}

\bibitem{serialization} Sergio Demian Lerner, {Buggy CVE-2013-4627 patch, open new vectors of attack}  \href{https://bitcointalk.org/index.php?topic=258770.0;prev_next=prev}{URL}

\bibitem{SPV} {Simplified Payment Verification}  \href{https://bitcoin.org/en/operating-modes-guide\#simplified-payment-verification-spv}{URL}, The Bitcoin protocol


\end{thebibliography}
\end{document}